\documentclass[pre, twocolumn, longbibliography, amsfonts, amssymb,
amsmath, nobalancelastpage, superscriptaddress]{revtex4-1}

\pdfoutput=1

\usepackage[OT1]{fontenc}
\usepackage{mathpazo,mathbbol,amsthm,bm}
\usepackage{graphicx}
\newcommand{\tool}{kruX}
\newcommand{\G}{\mathbf{G}}
\newcommand{\D}{\mathbf{D}}
\newcommand{\R}{\mathbf{R}}
\renewcommand{\S}{\mathbf{S}}
\newcommand{\I}{\mathbf{I}}
\newcommand{\N}{\mathbf{N}}
\newcommand{\Z}{\mathbf{Z}}
\newcommand{\K}{\mathbf{K}}

\begin{document}

\title{kruX: Matrix-based non-parametric eQTL discovery}

\author{Jianlong Qi}

\affiliation{School of Life Sciences -- LifeNet, Freiburg Institute
  for Advanced Studies (FRIAS), University of Freiburg, Albertstrasse
  19, D-79104 Freiburg im Breisgau, Germany} 

\affiliation{Epigenomic Mapping Centre, McGill University, Montreal,
  Quebec, Canada H3A0G1}

\author{Hassan Foroughi Asl}

\affiliation{Cardiovascular Genomics Group, Division of Vascular
  Biology, Department of Medical Biochemistry and Biophysics,
  Karolinska Institute, 171 77 Stockholm, Sweden}

\author{Johan Bj\"orkegren}

\affiliation{Cardiovascular Genomics Group, Division of Vascular
  Biology, Department of Medical Biochemistry and Biophysics,
  Karolinska Institute, 171 77 Stockholm, Sweden}

\affiliation{Department of Medical Pathology and Forensic Medicine,
  University of Tartu, 50090 Tartu, Estonia}

\author{Tom Michoel}

\email[Corresponding author, email:]{tom.michoel@roslin.ed.ac.uk}

\affiliation{School of Life Sciences -- LifeNet, Freiburg Institute
  for Advanced Studies (FRIAS), University of Freiburg, Albertstrasse
  19, D-79104 Freiburg im Breisgau, Germany}

\affiliation{The Roslin Institute, The University of Edinburgh, Easter
  Bush, Midlothian, EH25 9RG, Scotland, UK}

\begin{abstract} 
  The Kruskal-Wallis test is a popular non-parametric statistical test
  for identifying expression quantitative trait loci (eQTLs) from
  genome-wide data due to its robustness against variations in the
  underlying genetic model and expression trait distribution, but
  testing billions of marker-trait combinations one-by-one can become
  computationally prohibitive.  We developed kruX, an algorithm
  implemented in Matlab, Python and R that uses matrix multiplications
  to simultaneously calculate the Kruskal-Wallis test statistic for
  several millions of marker-trait combinations at once. KruX is more
  than ten thousand times faster than computing associations
  one-by-one on a typical human dataset. We used kruX and a dataset of
  more than 500k SNPs and 20k expression traits measured in 102 human
  blood samples to compare eQTLs detected by the Kruskal-Wallis test
  to eQTLs detected by the parametric ANOVA and linear model
  methods. We found that the Kruskal-Wallis test is more robust
  against data outliers and heterogeneous genotype group sizes and
  detects a higher proportion of non-linear associations, but is more
  conservative for calling additive linear associations. In summary,
  \tool{} enables the use of robust non-parametric methods for massive
  eQTL mapping without the need for a high-performance computing
  infrastructure and is freely available from
  \href{http://krux.googlecode.com}{http://krux.googlecode.com}.
\end{abstract}

\maketitle

\section{Background}

Genome-wide association studies have identified hundreds of DNA
variants associated to complex traits including disease in human alone
\cite{hindorff2009potential}. To understand how these variants affect
disease risk, genotype and organismal phenotype data are integrated
with intermediate molecular phenotypes to reconstruct disease networks
\cite{schadt2009}. A first step in this procedure is to identify DNA
variants that underpin variations in expression levels (eQTLs) of
transcripts \cite{cookson2009mapping}, proteins \cite{foss2007} or
metabolites \cite{nicholson2011genome}. As modern technologies
routinely produce genotype and expression data for a million or more
single-nucleotide polymorphisms (SNPs) and ten-thousands of molecular
abundance traits in a single experiment, often repeated across
multiple cell or tissue types, the number of statistical tests to be
performed when testing each SNP for association to each trait is
huge. Furthermore, multiple testing correction requires all tests to
be repeated several times on permuted data to generate an empirical
null distribution. Despite being trivially parallelisable, the
computational burden of testing SNP-trait associations one-by-one
quickly becomes prohibitive. 

Recently a new approach (``matrix-eQTL'') was developed which uses the
fact that the test statistics for the additive linear regression and
ANOVA models can be expressed as multiplications between rescaled
genotype and expression data matrices, thereby realising a dramatic
speed-up compared to traditional QTL-mapping algorithms
\cite{shabalin2012matrix}. A limitation of these models is their
assumption that the expression data is always normally distributed
within each genotype group. For this reason, QTL and eQTL studies have
frequently used non-parametric methods which are more robust against
variations in the underlying genetic model and trait distribution
\cite{kruglyak1995nonparametric,schadt2008}. In particular, the
non-parametric Kruskal-Wallis one-way analysis of variance
\cite{kruskal1952use} does not assume normal distributions and reports
small $P$-values if the median of at least one genotype group is
significantly different from the others \cite{schadt2008}.  

Here we report a matrix-based algorithm (``\tool{}''), implemented in
Matlab, Python and R, to simultaneously calculate the Kruskal-Wallis
test statistics for several millions of SNP-trait pairs at once that
is more than ten thousand times faster than calculating them one-by-one on a
human test dataset with more than 500,000 SNPs and 20,000 expression
traits. Additional benefits of \tool{} include the explicit handling
of missing values in both genotype and expression data and the support
of genetic markers with any number of alleles, including variable
allele numbers within a single dataset.

\section{Implementation}

\subsection{Input data}

KruX takes as input genotype values of $M$ genetic markers and
expression levels of $N$ transcripts, proteins or metabolites in $K$
individuals, organised in an $M\times K$ genotype matrix $\G$ and
$N\times K$ expression data matrix $\D$. Genetic markers take values
$0,1,\dots,\ell$, where $\ell$ is the maximum number of alleles
($\ell=2$ for biallelic markers), while molecular traits take
continuous values. We use built-in functions of Matlab, Python and R
to convert the expression data matrix $\D$ to a matrix $\R$ of data
ranks, ranked independently over each row (i.e. molecular trait). KruX
assumes that the input expression data has been adjusted for
covariates if it is necessary to do so
\cite{leek2007capturing,listgarten2010correction} and all data quality
control has been performed.

\subsection{Calculation of the Kruskal-Wallis test statistic by
   matrix multiplication}

 The genotype matrix $\G$ is first converted to sparse logical index
 matrices $\I_i$ of the same size, where $\I_i(m,k) = 1$ if $\G(m,k) =
 i$ and $0$ otherwise ($i=0,\dots,\ell$).  Next observe that the
 $1\times M$ vector $\N_i$ with entries $\N_i(m) = \sum_{k=1}^K
 \I_i(m,k)$ and $N\times M$ matrices $\S_i$ with entries
\begin{equation}\label{eq:4}
  \S_i(n,m) = \sum_{k=1}^K \R(n,k)\, \I_i(m,k)  = \bigl(\R\cdot \I_i^T\bigr)(n,m),
\end{equation}
are respectively the number of individuals and the sum of ranks for
the $n$th trait in the $i$th genotype group of the $m$th marker. We
can then calculate an $N\times M$ matrix $\S$ with entries
\begin{equation}\label{eq:2}
  \S(n,m) = \frac{12}{K(K+1)} \sum_{i=0}^\ell \frac{\S_i(n,m)^2}{\N_i(m)} - 3(K+1),
\end{equation}
using efficient vectorised operations.  If none of the rows in $\D$
contain ties, then each entry $\S(n,m)$ equals the Kruskal-Wallis test
statistic for testing trait $n$ against marker $m$
\cite{kruskal1952use}.  For markers with less than the maximum of
$\ell$ genotype values, $0/0$ division will result in NaN columns in
the intermediate matrices with entries $\S_i(n,m)^2/\N_i(m)$ for the
empty genotype groups. By replacing all NaN's by zeros before making
the sum in eq. (\ref{eq:2}), the corresponding entries in $\S$ will be
the correct statistics for a test with fewer than $\ell$ degrees of
freedom.  Thus we need $\ell+1$ matrix multiplications and the
associated element-wise operations to calculate the test statistic
values for all marker-trait combinations.

\subsection{P-value calculation and empirical FDR correction}

KruX takes as input a $P$-value threshold $P_c$, calculates the
corresponding test statistic thresholds for $d$ degrees of freedom
($d=1,\dots,\ell-1$), and identifies the entries in $S$ which exceed
the appropriate threshold value. For these entries only a $P$-value is
calculated using the $\chi^2$ distribution.  Empirical false-discovery
rate (FDR) values are computed by repeating the $P$-value calculation
(with the same $P_c$) multiple times on data where the columns of the
expression data ranks are randomly permuted. The FDR value for any
value $P\leq P_c$ is defined as the ratio of the average number of
associations with $P'\leq P$ in the randomised data to the number of
associations with $P'\leq P$ in the real data.

\subsection{Handling missing values}

When data values are missing for some marker or trait, all test
statistics for that marker or trait need to be adjusted for a smaller
number of observations. For the expression data, missing values are
easily handled since the ranking algorithms will give NaN's the
highest rank. By setting the entries corresponding to missing values
in $\D$ to zero in $\R$, eq. (\ref{eq:4}) still produces the correct
sums of ranks, while the matrix multiplication
\begin{align*}
  \bigl(\Z\cdot \I_i^T\bigr)(n,m) = \sum_{k=1}^K \Z(n,k)\,  I_i(m,k) &= \N_i(n,m),
\end{align*}
where $\Z$ is the $N\times K$ matrix with $\Z(n,k)=0$ whenever
$\D(n,k)=\text{NaN}$ and $1$ otherwise, produces the corrected number
of individuals in the $i$th group of the $m$th marker for the $n$th
trait. Replacing the constant $K$ in eq. (\ref{eq:2}) by a $N\times M$
matrix $\K$ where $\K(n,m)$ is the number of non-missing samples for trait
$n$ and performing element-wise division and substraction operations
then gives the correct test statistic for all pairs.

Handling missing genotype data is less easy because the expression
ranks that need to be adjusted are specific to each marker-trait
combination (e.g if a marker has a missing value where a trait has
rank $r_1$, then all samples with ranks $r=r_1+1,\dots,K$ need to be
lowered by $1$). KruX uses the fact that missing genotype values are
generally due to sample quality and therefore patterns of missing
values are often repeated among markers. For each unique missing value
pattern, a new genotype matrix for all markers with that pattern and a
new expression data matrix with the corresponding samples removed are
constructed to calculate the test statistics for all affected marker
gene combinations. Missing genotype data increases the computational
cost of the algorithm considerably and it is recommended to limit the
number of missing values by only considering markers with a
sufficiently high call rate.

\subsection{Handling tied data}

In the presence of tied observations, the statistic in
eq. \eqref{eq:2} needs to be divided by a factor $1 - \frac{\sum
  T}{K^3-K}$, where the summation is over all groups of ties and
$T=t^3-t$ for each group of ties, with $t$ the number of tied data in
the group \cite{kruskal1952use}. The factor $T$ is automatically
computed for each trait during the ranking step and the matrix $S$ is
therefore easily corrected using element-wise matrix operations
(Matlab version only). Whereas ties are usually rare in standard gene
expression datasets, the ability to handle tied data expands the scope
of \tool{} to count-based, discretised or qualitative data types.

\subsection{Data slicing}

Since \tool{} needs to create intermediate matrices of size $N\times
M$, where $N$ is the number of traits and $M$ the number of markers,
which do not usually fit into memory for large datasets, \tool{} supports the
use of data `slices' to divide the complete data into manageable
chunks. In typical applications, the number of markers is one or two
orders of magnitude larger than the number of traits. Therefore the
default behaviour of \tool{} is to keep the expression data as a
single matrix and simultaneously test all traits against subsets of
markers. The user can provide either a slice size and \tool{} will
process marker blocks of this size serially, or a slice size and
initial marker and \tool{} will process a single slice starting from
that marker. The latter option allows trivial parallelisation across
multiple processors.

\section{Results and Discussion}

\subsection{Validation data}

To test kruX we provide example analysis scripts and a small
anonymised dataset of 2,000 randomly selected genes and markers from
100 randomly selected yeast segregants \cite{brem2005landscape}. Here
we describe an application of kruX on a human dataset of 19,610 genes
and 530,222 SNP markers measured in 102 whole blood samples from the
Stockholm Atherosclerosis Gene Expression (STAGE) study
\cite{hagg2009multi}. All SNPs in the dataset had minor allele
frequency greater than $5\%$, no missing values and probability to be
in Hardy-Weinberg equilibrium greater than $10^{-6}$.

\subsection{kruX is exact and fast}

\begin{figure}[t]
  \includegraphics[width=\linewidth]{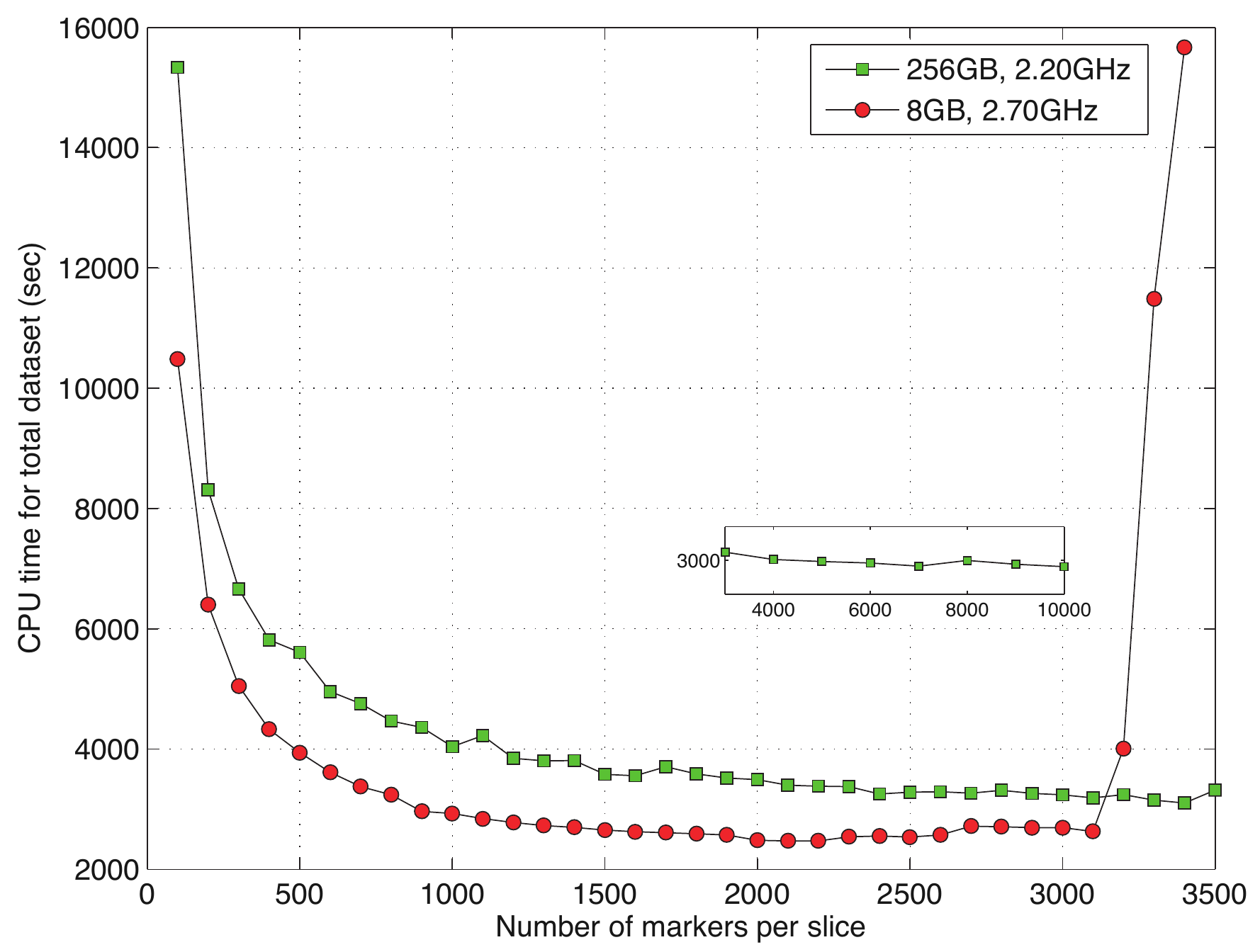}
  \caption{\textbf{kruX runtime on STAGE data.} Total extrapolated
    single-CPU run time in seconds for the Matlab implementation of
    kruX for different numbers of SNP markers per data slice (see main
    text for details). Green squares are times on a high-memory server
    with 256 GB RAM and 2.20 GHz processor and red circles are times
    on a laptop with 8 GB RAM and 2.70 GHz processor. The insert shows
    the continuation of the green squares upto a slice size of 10,000
    markers.}
  \label{fig:cpu}
\end{figure}

We first confirmed that \tool{} produces the same results as testing
marker-trait combinations one-by-one using the built-in Kruskal-Wallis
functions to verify the correctness of our implementations.  To test
the performance of \tool{} we divided the genotype data into slices of
variable size and extrapolated the total run time from running a
single genotype data slice against all expression traits and
multiplying by the number of slices needed to cover the entire set of
530,222 SNPs. The total run time rapidly decreases until a genotype
slice contains about 1,000 SNPs and stays almost constant
thereafter. On a laptop with 8 GB RAM, the limit is reached at around
3,000 SNPs per slice after which run time sharply increases again due
to memory limitations (Fig.~\ref{fig:cpu}). We therefore recommend a
genotype slice size of around 2,000 markers, resulting for this
dataset in around 250 separate jobs, which will take around 2,500
seconds (42 minutes) when run serially on a single processor.  By
comparison, the total extrapolated run time when computing all 19,610
$\times$ 530,222 associations one-by-one using the built-in
Kruskal-Wallis function on the same hardware as in Fig.~\ref{fig:cpu}
are respectively $4.8\cdot 10^7$ (256 GB, 2.20 GHz server) and
$2.6\cdot 10^7$ (8 GB, 2.70 GHz laptop) seconds such that kruX is
respectively 17,000 and 11,000 times faster on this particular
dataset. On the same dataset and hardware, the comparatively simpler
matrix operations for the parametric tests in matrix-eQTL took
respectively 5 minutes (linear model) and 7.4 minutes (ANOVA model).

\subsection{The Kruskal-Wallis test is more conservative than
  corresponding parametric tests}

\begin{figure}[t]
  \includegraphics[width=\linewidth]{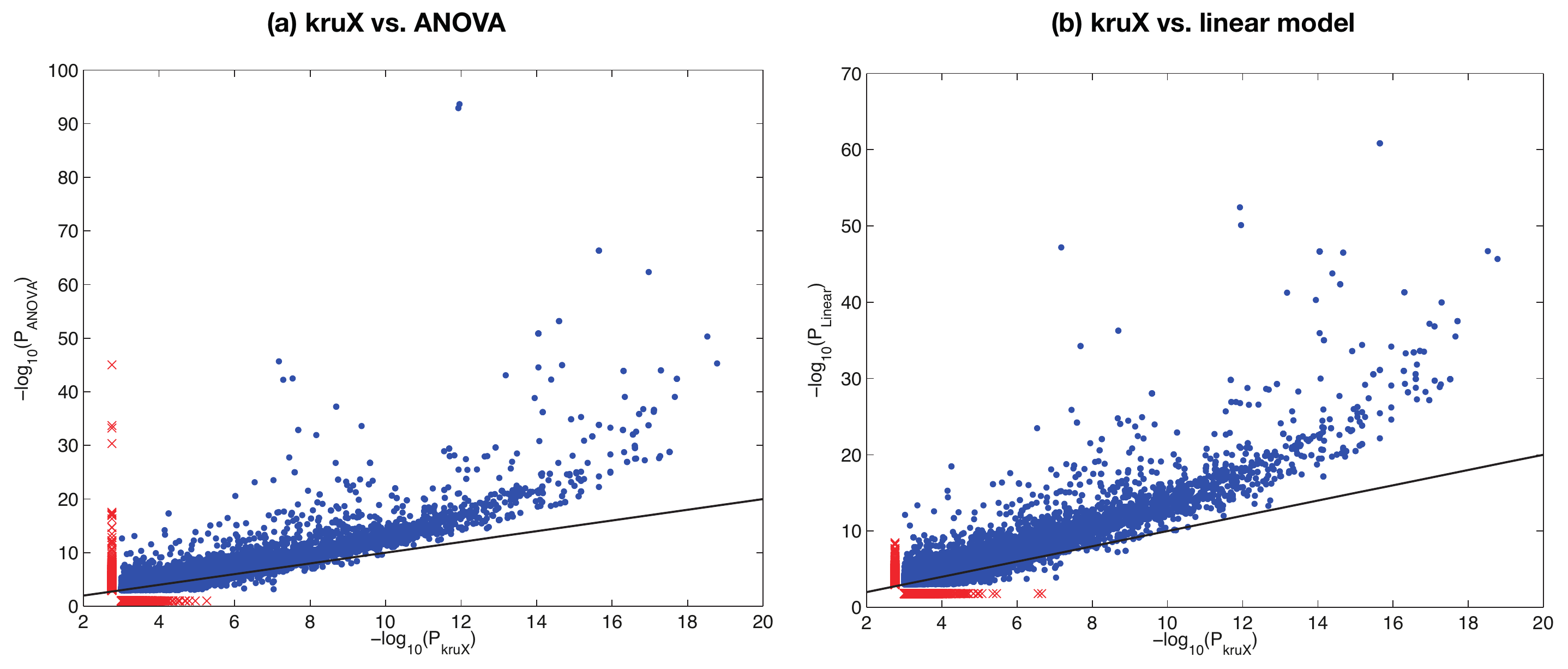}
  \caption{\textbf{Comparison of kruX vs. parametric ANOVA and
      linear models.} Comparison of nominal non-parametric $P$-values
    calculated by kruX vs. parametric ANOVA \textbf{(a)} and linear
    models \textbf{(b)}, showing all \textit{cis}-acting eQTL-gene
    pairs with $P<10^{-3}$ detected by both methods (blue dots) and by
    only one of the methods (red crosses). The black line indicates
    the line with slope $y=x$.}
  \label{fig:pval}
\end{figure}

\begin{figure}[t]
  \includegraphics[width=\linewidth]{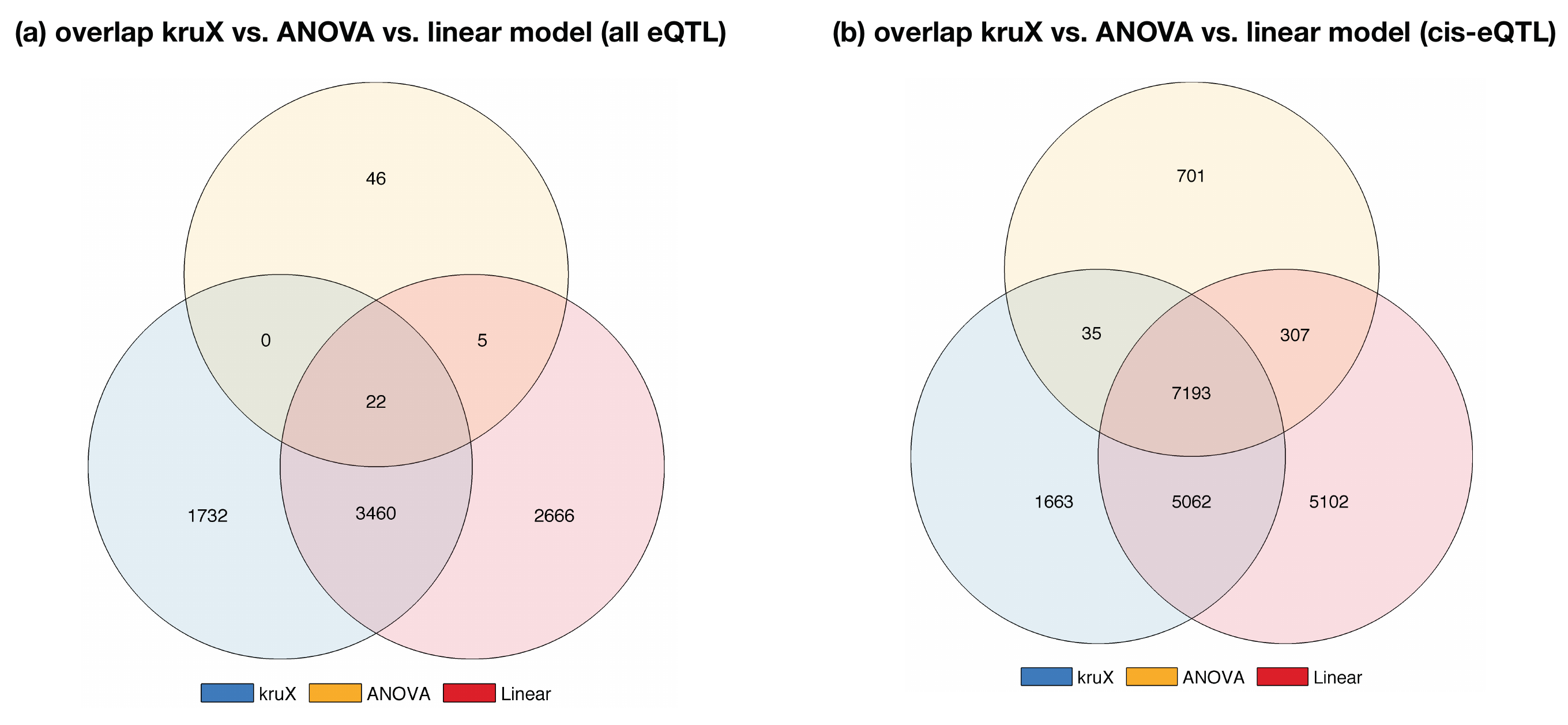}
  \caption{\textbf{Comparison of kruX vs. parametric ANOVA and
      linear models.} Comparison of all eQTL-gene pairs (FDR=30\%)
    \textbf{(a)} and all \textit{cis}-acting eQTL-gene pairs
    (FDR=10\%) \textbf{(b)} after empirical FDR correction between
    kruX (blue lower left set), parametric ANOVA (yellow upper set),
    and linear models (red lower right set).}
  \label{fig:fdr}
\end{figure}

Next we compared the output of \tool{} and matrix-eQTL's parametric
ANOVA and linear model (henceforth called ``ANOVA'' and ``linear'')
methods. The Kruskal-Wallis test is more conservative than the ANOVA
and linear methods, i.e. it has a higher nominal $P$-value for almost
all marker-trait combinations (Fig.~\ref{fig:pval}). 
Since random data
will be subjected to the same biases, nominal $P$-values cannot be
directly compared to assess significance. We therefore performed
empirical FDR correction for multiple testing using three randomised
datasets (cf. Implementation).  Surprisingly, after FDR correction only a
limited number of associations remained for ANOVA even at an FDR
threshold of $30\%$, whereas the number of associations detected by
kruX and the linear method was comparable (Fig.~\ref{fig:fdr}(a)).
Detailed analysis showed that this is due to pairing of SNPs with rare
homozygous minor alleles (one or two samples) to genes with outlier
expression levels, resulting in extremely low $P$-values for the ANOVA
method in real as well as randomised data (see also below).  To reduce
the incidence of chance associations between singleton genotype groups
and outlying expression values in the ANOVA method we repeated the
empirical FDR correction, this time keeping only marker-trait
combinations within 1Mbp of each other (``cis-eQTLs''). At an FDR
threshold of $10\%$ the number of significant \textit{cis}-eQTL-gene
pairs is indeed comparable between the three methods, with a large
proportion of pairs detected by all three of them
(Fig.~\ref{fig:fdr}(b)).

\subsection{The Kruskal-Wallis test is more robust and detects more
  non-linear associations}

\begin{figure}[t]
  \includegraphics[width=\linewidth]{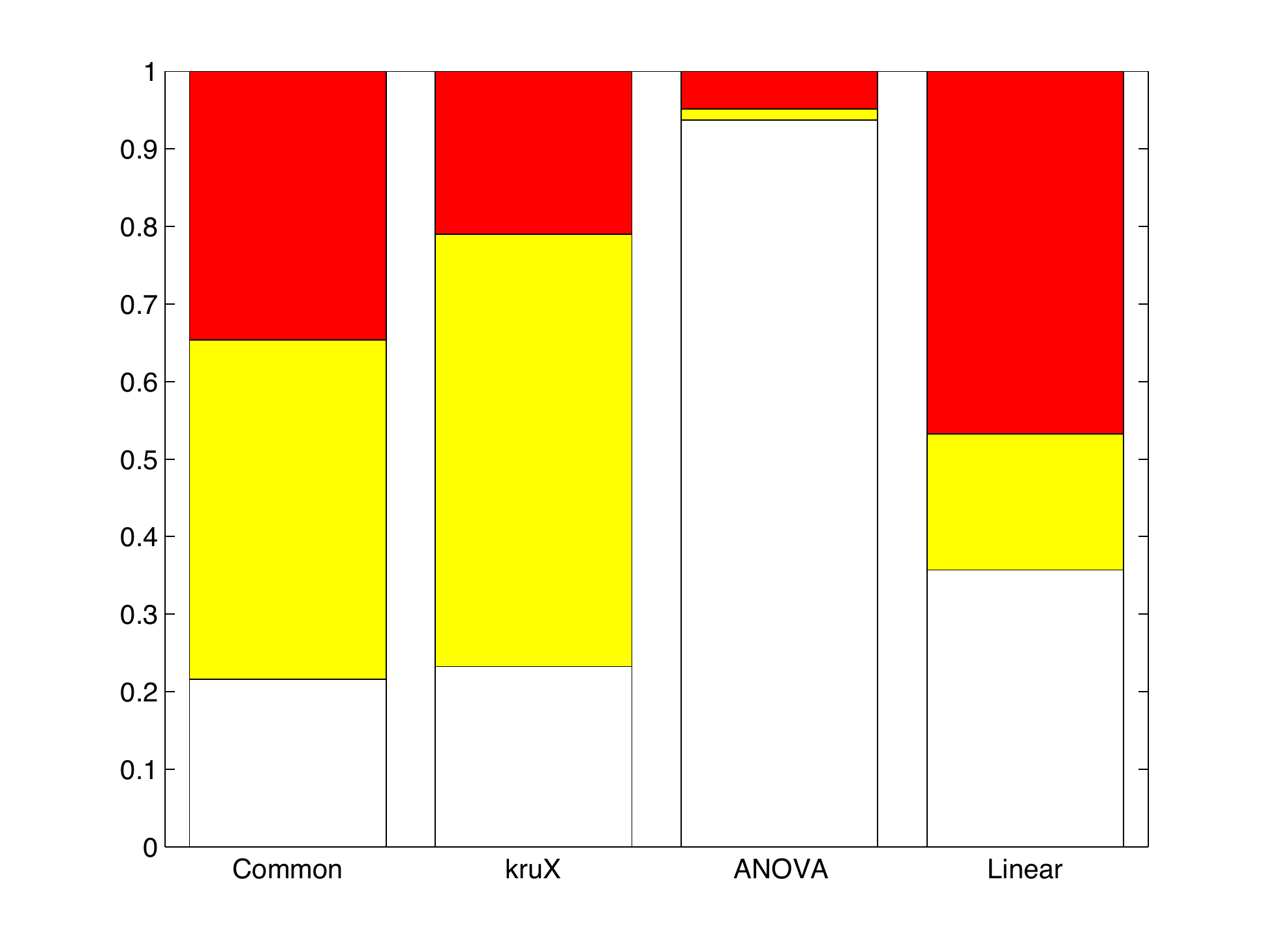}
  \caption{\textbf{Relative proportions of eQTL types.}  Relative
    proportion of eQTL-types for cis-eQTLs common to all 3 methods and
    specific to each method; white (bottom), skewed genotype group
    sizes; yellow (middle), non-linear eQTLs; red (top), others. The
    absolute number of eQTLs in each group is 7,193 (Common), 1,663
    (kruX), 701 (ANOVA) and 5,102 (Linear), cf. Fig.~\ref{fig:fdr}(b).}
  \label{fig:groups}
\end{figure}

\begin{figure}[t]
  \includegraphics[width=\linewidth]{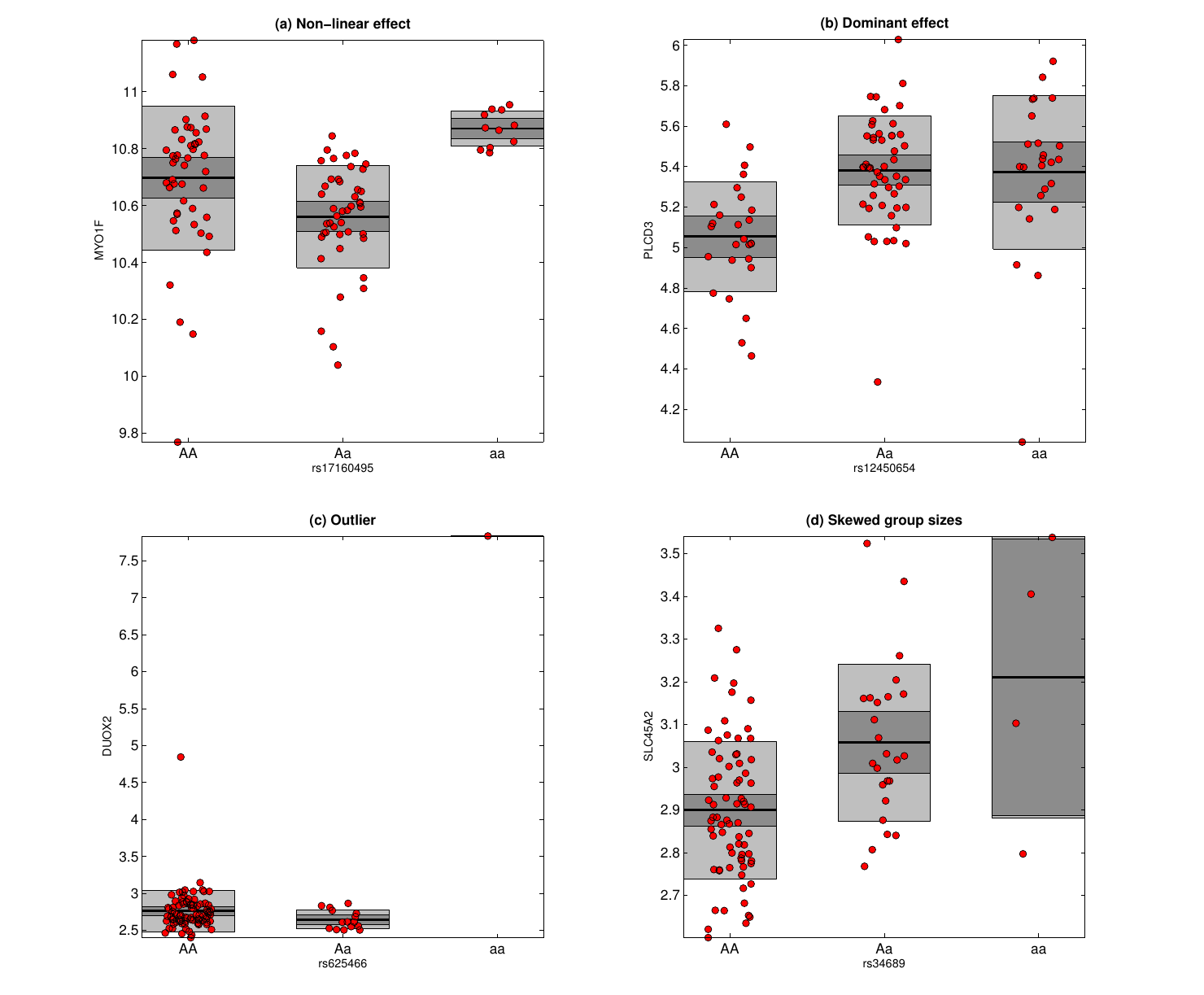}
  \caption{\textbf{Representative examples of eQTL associations.}
    \textbf{(a-b) Non-linear associations.} kruX identifies more
    non-linear relations where the gene expression level of the
    heterozygous samples lies outside the typical range of the
    homozygous samples \textbf{(a)} or where one allele has a dominant
    effect on the gene expression level \textbf{(b)}. \textbf{(c-d)
      Problematic associations.}  Parametric ANOVA gives high
    significance to spurious associations for genes with outlying
    expression samples that coincide with singleton genotype groups
    \textbf{(c)}.  Associations with skewed genotype group sizes where
    the model assumptions are difficult to ascertain achieve high
    significance using linear models \textbf{(d)}.}
  \label{fig:box}
\end{figure}

We classified eQTL-gene pairs as ``skewed group sizes'' (smallest
genotype group less than 5 elements), non-skewed ``non-linear''
[median of heterozygous and homozygous samples significantly different
(Wilcoxon rank sum $P<0.05$)] and non-skewed ``other'' (all
others). \textit{Cis}-associations identified exclusively by the
Kruskal-Wallis test are more often non-linear and the overall
distribution of eQTL-types is more similar to associations identified
by all three methods, compared to the ANOVA and linear methods
(Fig.~\ref{fig:groups} and Fig.~\ref{fig:box}(a-b)). Of the 701
associations exclusively identified using the parametric ANOVA method,
657 ($94\%$) had skewed group sizes, including 426 ($61\%$) with a
singleton genotype group (the aforementioned `outliers',
cf. Fig.~\ref{fig:box}(c)). The associations exclusively identified by
the linear method also contained a much higher proportion of SNPs with
skewed group sizes than the corresponding kruX associations ($36\%$
vs. $23\%$) and, as expected, a reduced number of non-linear
associations (Fig.~\ref{fig:groups} and Fig.~\ref{fig:box}(d)).

\section{Conclusions}

We have developed kruX, a software tool that uses matrix
multiplications to simultaneously calculate the Kruskal-Wallis test
statistics for millions of marker-trait combinations in a single
operation, thereby realising a dramatic speed-up compared to
calculating the test statistics one-by-one.  The availability of a
fast method to identify eQTL associations using a non-parametric test
allowed us to assess in more detail how differences in model
assumptions compared to parametric methods lead to differences in
identified eQTLs. Our results on a typical human dataset indicate that
the the parametric ANOVA method is highly sensitive to the presence of
outlying gene expression values and SNPs with singleton genotype
groups. We caution against its use without prior filtering of such
outliers. Linear models reported the highest number of eQTL
associations after empirical FDR correction. These are understandably
biased towards additive linear associations and were also sensitive to
the presence of skewed genotype group sizes, albeit to a much lesser
extent than the parametric ANOVA method. The Kruskal-Wallis test on
the other hand is robust against data outliers and heterogeneous
genotype group sizes and detects a higher proportion of non-linear
associations, but it is more conservative for calling additive linear
associations than linear models, even after FDR correction.

In summary, \tool{} enables the use of non-parametric methods for
massive eQTL mapping without the need for a high-performance computing
infrastructure.

\begin{acknowledgements}
  This work was supported by the Freiburg Institute for Advanced
  Studies and Roslin Institute Strategic Grant funding from the BBSRC.
\end{acknowledgements}


\begin{thebibliography}{13}%
\makeatletter
\providecommand \@ifxundefined [1]{%
 \@ifx{#1\undefined}
}%
\providecommand \@ifnum [1]{%
 \ifnum #1\expandafter \@firstoftwo
 \else \expandafter \@secondoftwo
 \fi
}%
\providecommand \@ifx [1]{%
 \ifx #1\expandafter \@firstoftwo
 \else \expandafter \@secondoftwo
 \fi
}%
\providecommand \natexlab [1]{#1}%
\providecommand \enquote  [1]{``#1''}%
\providecommand \bibnamefont  [1]{#1}%
\providecommand \bibfnamefont [1]{#1}%
\providecommand \citenamefont [1]{#1}%
\providecommand \href@noop [0]{\@secondoftwo}%
\providecommand \href [0]{\begingroup \@sanitize@url \@href}%
\providecommand \@href[1]{\@@startlink{#1}\@@href}%
\providecommand \@@href[1]{\endgroup#1\@@endlink}%
\providecommand \@sanitize@url [0]{\catcode `\\12\catcode `\$12\catcode
  `\&12\catcode `\#12\catcode `\^12\catcode `\_12\catcode `\%12\relax}%
\providecommand \@@startlink[1]{}%
\providecommand \@@endlink[0]{}%
\providecommand \url  [0]{\begingroup\@sanitize@url \@url }%
\providecommand \@url [1]{\endgroup\@href {#1}{\urlprefix }}%
\providecommand \urlprefix  [0]{URL }%
\providecommand \Eprint [0]{\href }%
\providecommand \doibase [0]{http://dx.doi.org/}%
\providecommand \selectlanguage [0]{\@gobble}%
\providecommand \bibinfo  [0]{\@secondoftwo}%
\providecommand \bibfield  [0]{\@secondoftwo}%
\providecommand \translation [1]{[#1]}%
\providecommand \BibitemOpen [0]{}%
\providecommand \bibitemStop [0]{}%
\providecommand \bibitemNoStop [0]{.\EOS\space}%
\providecommand \EOS [0]{\spacefactor3000\relax}%
\providecommand \BibitemShut  [1]{\csname bibitem#1\endcsname}%
\let\auto@bib@innerbib\@empty
\bibitem [{\citenamefont {Hindorff}\ \emph {et~al.}(2009)\citenamefont
  {Hindorff}, \citenamefont {Sethupathy}, \citenamefont {Junkins},
  \citenamefont {Ramos}, \citenamefont {Mehta}, \citenamefont {Collins},\ and\
  \citenamefont {Manolio}}]{hindorff2009potential}%
  \BibitemOpen
  \bibfield  {author} {\bibinfo {author} {\bibfnamefont {Lucia~A}\ \bibnamefont
  {Hindorff}}, \bibinfo {author} {\bibfnamefont {Praveen}\ \bibnamefont
  {Sethupathy}}, \bibinfo {author} {\bibfnamefont {Heather~A}\ \bibnamefont
  {Junkins}}, \bibinfo {author} {\bibfnamefont {Erin~M}\ \bibnamefont {Ramos}},
  \bibinfo {author} {\bibfnamefont {Jayashri~P}\ \bibnamefont {Mehta}},
  \bibinfo {author} {\bibfnamefont {Francis~S}\ \bibnamefont {Collins}}, \ and\
  \bibinfo {author} {\bibfnamefont {Teri~A}\ \bibnamefont {Manolio}},\
  }\bibfield  {title} {\enquote {\bibinfo {title} {Potential etiologic and
  functional implications of genome-wide association loci for human diseases
  and traits},}\ }\href@noop {} {\bibfield  {journal} {\bibinfo  {journal}
  {Proceedings of the National Academy of Sciences}\ }\textbf {\bibinfo
  {volume} {106}},\ \bibinfo {pages} {9362--9367} (\bibinfo {year}
  {2009})}\BibitemShut {NoStop}%
\bibitem [{\citenamefont {Schadt}(2009)}]{schadt2009}%
  \BibitemOpen
  \bibfield  {author} {\bibinfo {author} {\bibfnamefont {E~E}\ \bibnamefont
  {Schadt}},\ }\bibfield  {title} {\enquote {\bibinfo {title} {Molecular
  networks as sensors and drivers of common human diseases},}\ }\href@noop {}
  {\bibfield  {journal} {\bibinfo  {journal} {Nature}\ }\textbf {\bibinfo
  {volume} {461}},\ \bibinfo {pages} {218--223} (\bibinfo {year}
  {2009})}\BibitemShut {NoStop}%
\bibitem [{\citenamefont {Cookson}\ \emph {et~al.}(2009)\citenamefont
  {Cookson}, \citenamefont {Liang}, \citenamefont {Abecasis}, \citenamefont
  {Moffatt},\ and\ \citenamefont {Lathrop}}]{cookson2009mapping}%
  \BibitemOpen
  \bibfield  {author} {\bibinfo {author} {\bibfnamefont {William}\ \bibnamefont
  {Cookson}}, \bibinfo {author} {\bibfnamefont {Liming}\ \bibnamefont {Liang}},
  \bibinfo {author} {\bibfnamefont {Gon{\c{c}}alo}\ \bibnamefont {Abecasis}},
  \bibinfo {author} {\bibfnamefont {Miriam}\ \bibnamefont {Moffatt}}, \ and\
  \bibinfo {author} {\bibfnamefont {Mark}\ \bibnamefont {Lathrop}},\ }\bibfield
   {title} {\enquote {\bibinfo {title} {Mapping complex disease traits with
  global gene expression},}\ }\href@noop {} {\bibfield  {journal} {\bibinfo
  {journal} {Nature Reviews Genetics}\ }\textbf {\bibinfo {volume} {10}},\
  \bibinfo {pages} {184--194} (\bibinfo {year} {2009})}\BibitemShut {NoStop}%
\bibitem [{\citenamefont {Foss}\ \emph {et~al.}(2007)\citenamefont {Foss},
  \citenamefont {Radulovic}, \citenamefont {Shaffer}, \citenamefont {Ruderfer},
  \citenamefont {Bedalov}, \citenamefont {Goodlett},\ and\ \citenamefont
  {Kruglyak}}]{foss2007}%
  \BibitemOpen
  \bibfield  {author} {\bibinfo {author} {\bibfnamefont {E~J}\ \bibnamefont
  {Foss}}, \bibinfo {author} {\bibfnamefont {D}~\bibnamefont {Radulovic}},
  \bibinfo {author} {\bibfnamefont {S~A}\ \bibnamefont {Shaffer}}, \bibinfo
  {author} {\bibfnamefont {D~M}\ \bibnamefont {Ruderfer}}, \bibinfo {author}
  {\bibfnamefont {A}~\bibnamefont {Bedalov}}, \bibinfo {author} {\bibfnamefont
  {D~R}\ \bibnamefont {Goodlett}}, \ and\ \bibinfo {author} {\bibfnamefont
  {L}~\bibnamefont {Kruglyak}},\ }\bibfield  {title} {\enquote {\bibinfo
  {title} {Gentic basis of proteome variation in yeast},}\ }\href@noop {}
  {\bibfield  {journal} {\bibinfo  {journal} {Nature Genetics}\ }\textbf
  {\bibinfo {volume} {39}},\ \bibinfo {pages} {1369--1375} (\bibinfo {year}
  {2007})}\BibitemShut {NoStop}%
\bibitem [{\citenamefont {Nicholson}\ \emph {et~al.}(2011)\citenamefont
  {Nicholson}, \citenamefont {Rantalainen}, \citenamefont {Li}, \citenamefont
  {Maher}, \citenamefont {Malmodin}, \citenamefont {Ahmadi}, \citenamefont
  {Faber}, \citenamefont {Barrett}, \citenamefont {Min}, \citenamefont {Rayner}
  \emph {et~al.}}]{nicholson2011genome}%
  \BibitemOpen
  \bibfield  {author} {\bibinfo {author} {\bibfnamefont {George}\ \bibnamefont
  {Nicholson}}, \bibinfo {author} {\bibfnamefont {Mattias}\ \bibnamefont
  {Rantalainen}}, \bibinfo {author} {\bibfnamefont {Jia~V}\ \bibnamefont {Li}},
  \bibinfo {author} {\bibfnamefont {Anthony~D}\ \bibnamefont {Maher}}, \bibinfo
  {author} {\bibfnamefont {Daniel}\ \bibnamefont {Malmodin}}, \bibinfo {author}
  {\bibfnamefont {Kourosh~R}\ \bibnamefont {Ahmadi}}, \bibinfo {author}
  {\bibfnamefont {Johan~H}\ \bibnamefont {Faber}}, \bibinfo {author}
  {\bibfnamefont {Amy}\ \bibnamefont {Barrett}}, \bibinfo {author}
  {\bibfnamefont {Josine~L}\ \bibnamefont {Min}}, \bibinfo {author}
  {\bibfnamefont {N~William}\ \bibnamefont {Rayner}},  \emph {et~al.},\
  }\bibfield  {title} {\enquote {\bibinfo {title} {A genome-wide metabolic
  {QTL} analysis in {E}uropeans implicates two loci shaped by recent positive
  selection},}\ }\href@noop {} {\bibfield  {journal} {\bibinfo  {journal} {PLoS
  Genetics}\ }\textbf {\bibinfo {volume} {7}},\ \bibinfo {pages} {e1002270}
  (\bibinfo {year} {2011})}\BibitemShut {NoStop}%
\bibitem [{\citenamefont {Shabalin}(2012)}]{shabalin2012matrix}%
  \BibitemOpen
  \bibfield  {author} {\bibinfo {author} {\bibfnamefont {Andrey~A}\
  \bibnamefont {Shabalin}},\ }\bibfield  {title} {\enquote {\bibinfo {title}
  {Matrix {eQTL}: ultra fast {eQTL} analysis via large matrix operations},}\
  }\href@noop {} {\bibfield  {journal} {\bibinfo  {journal} {Bioinformatics}\
  }\textbf {\bibinfo {volume} {28}},\ \bibinfo {pages} {1353--1358} (\bibinfo
  {year} {2012})}\BibitemShut {NoStop}%
\bibitem [{\citenamefont {Kruglyak}\ and\ \citenamefont
  {Lander}(1995)}]{kruglyak1995nonparametric}%
  \BibitemOpen
  \bibfield  {author} {\bibinfo {author} {\bibfnamefont {Leonid}\ \bibnamefont
  {Kruglyak}}\ and\ \bibinfo {author} {\bibfnamefont {Eric~S}\ \bibnamefont
  {Lander}},\ }\bibfield  {title} {\enquote {\bibinfo {title} {A nonparametric
  approach for mapping quantitative trait loci},}\ }\href@noop {} {\bibfield
  {journal} {\bibinfo  {journal} {Genetics}\ }\textbf {\bibinfo {volume}
  {139}},\ \bibinfo {pages} {1421--1428} (\bibinfo {year} {1995})}\BibitemShut
  {NoStop}%
\bibitem [{\citenamefont {Schadt}\ \emph {et~al.}(2008)\citenamefont {Schadt},
  \citenamefont {Molony}, \citenamefont {Chudin}, \citenamefont {Hao},
  \citenamefont {Yang}, \citenamefont {Lum}, \citenamefont {Kasarskis},
  \citenamefont {Zhang}, \citenamefont {Wang}, \citenamefont {Suver},
  \citenamefont {Zhu}, \citenamefont {Millstein}, \citenamefont {Sieberts},
  \citenamefont {Lamb}, \citenamefont {GuhaThakurta}, \citenamefont {Derry},
  \citenamefont {Storey}, \citenamefont {Avila-Campillo}, \citenamefont
  {Kruger}, \citenamefont {Johnson}, \citenamefont {Rohl}, \citenamefont {van
  Nas}, \citenamefont {Mehrabian}, \citenamefont {Drake}, \citenamefont
  {Lusis}, \citenamefont {Smith}, \citenamefont {Guengerich}, \citenamefont
  {Strom}, \citenamefont {Schuetz}, \citenamefont {Rushmore},\ and\
  \citenamefont {Ulrich}}]{schadt2008}%
  \BibitemOpen
  \bibfield  {author} {\bibinfo {author} {\bibfnamefont {E.~E.}\ \bibnamefont
  {Schadt}}, \bibinfo {author} {\bibfnamefont {C.}~\bibnamefont {Molony}},
  \bibinfo {author} {\bibfnamefont {E.}~\bibnamefont {Chudin}}, \bibinfo
  {author} {\bibfnamefont {K.}~\bibnamefont {Hao}}, \bibinfo {author}
  {\bibfnamefont {X.}~\bibnamefont {Yang}}, \bibinfo {author} {\bibfnamefont
  {P.~Y.}\ \bibnamefont {Lum}}, \bibinfo {author} {\bibfnamefont
  {A.}~\bibnamefont {Kasarskis}}, \bibinfo {author} {\bibfnamefont
  {B.}~\bibnamefont {Zhang}}, \bibinfo {author} {\bibfnamefont
  {S.}~\bibnamefont {Wang}}, \bibinfo {author} {\bibfnamefont {C.}~\bibnamefont
  {Suver}}, \bibinfo {author} {\bibfnamefont {J.}~\bibnamefont {Zhu}}, \bibinfo
  {author} {\bibfnamefont {J.}~\bibnamefont {Millstein}}, \bibinfo {author}
  {\bibfnamefont {S.}~\bibnamefont {Sieberts}}, \bibinfo {author}
  {\bibfnamefont {J.}~\bibnamefont {Lamb}}, \bibinfo {author} {\bibfnamefont
  {D.}~\bibnamefont {GuhaThakurta}}, \bibinfo {author} {\bibfnamefont
  {J.}~\bibnamefont {Derry}}, \bibinfo {author} {\bibfnamefont {J.~D.}\
  \bibnamefont {Storey}}, \bibinfo {author} {\bibfnamefont {I.}~\bibnamefont
  {Avila-Campillo}}, \bibinfo {author} {\bibfnamefont {M.~J.}\ \bibnamefont
  {Kruger}}, \bibinfo {author} {\bibfnamefont {J.~M.}\ \bibnamefont {Johnson}},
  \bibinfo {author} {\bibfnamefont {C.~A.}\ \bibnamefont {Rohl}}, \bibinfo
  {author} {\bibfnamefont {A.}~\bibnamefont {van Nas}}, \bibinfo {author}
  {\bibfnamefont {M.}~\bibnamefont {Mehrabian}}, \bibinfo {author}
  {\bibfnamefont {T.~A.}\ \bibnamefont {Drake}}, \bibinfo {author}
  {\bibfnamefont {A.~J.}\ \bibnamefont {Lusis}}, \bibinfo {author}
  {\bibfnamefont {R.~C.}\ \bibnamefont {Smith}}, \bibinfo {author}
  {\bibfnamefont {F.~P.}\ \bibnamefont {Guengerich}}, \bibinfo {author}
  {\bibfnamefont {S.~C.}\ \bibnamefont {Strom}}, \bibinfo {author}
  {\bibfnamefont {E.}~\bibnamefont {Schuetz}}, \bibinfo {author} {\bibfnamefont
  {T.~H.}\ \bibnamefont {Rushmore}}, \ and\ \bibinfo {author} {\bibfnamefont
  {R.}~\bibnamefont {Ulrich}},\ }\bibfield  {title} {\enquote {\bibinfo {title}
  {{{M}apping the genetic architecture of gene expression in human liver}},}\
  }\href@noop {} {\bibfield  {journal} {\bibinfo  {journal} {PLoS Biol.}\
  }\textbf {\bibinfo {volume} {6}},\ \bibinfo {pages} {e107} (\bibinfo {year}
  {2008})}\BibitemShut {NoStop}%
\bibitem [{\citenamefont {Kruskal}\ and\ \citenamefont
  {Wallis}(1952)}]{kruskal1952use}%
  \BibitemOpen
  \bibfield  {author} {\bibinfo {author} {\bibfnamefont {William~H}\
  \bibnamefont {Kruskal}}\ and\ \bibinfo {author} {\bibfnamefont {W~Allen}\
  \bibnamefont {Wallis}},\ }\bibfield  {title} {\enquote {\bibinfo {title} {Use
  of ranks in one-criterion variance analysis},}\ }\href@noop {} {\bibfield
  {journal} {\bibinfo  {journal} {Journal of the American Statistical
  Association}\ }\textbf {\bibinfo {volume} {47}},\ \bibinfo {pages} {583--621}
  (\bibinfo {year} {1952})}\BibitemShut {NoStop}%
\bibitem [{\citenamefont {Leek}\ and\ \citenamefont
  {Storey}(2007)}]{leek2007capturing}%
  \BibitemOpen
  \bibfield  {author} {\bibinfo {author} {\bibfnamefont {Jeffrey~T}\
  \bibnamefont {Leek}}\ and\ \bibinfo {author} {\bibfnamefont {John~D}\
  \bibnamefont {Storey}},\ }\bibfield  {title} {\enquote {\bibinfo {title}
  {Capturing heterogeneity in gene expression studies by surrogate variable
  analysis},}\ }\href@noop {} {\bibfield  {journal} {\bibinfo  {journal} {PLoS
  Genetics}\ }\textbf {\bibinfo {volume} {3}},\ \bibinfo {pages} {e161}
  (\bibinfo {year} {2007})}\BibitemShut {NoStop}%
\bibitem [{\citenamefont {Listgarten}\ \emph {et~al.}(2010)\citenamefont
  {Listgarten}, \citenamefont {Kadie}, \citenamefont {Schadt},\ and\
  \citenamefont {Heckerman}}]{listgarten2010correction}%
  \BibitemOpen
  \bibfield  {author} {\bibinfo {author} {\bibfnamefont {Jennifer}\
  \bibnamefont {Listgarten}}, \bibinfo {author} {\bibfnamefont {Carl}\
  \bibnamefont {Kadie}}, \bibinfo {author} {\bibfnamefont {Eric~E}\
  \bibnamefont {Schadt}}, \ and\ \bibinfo {author} {\bibfnamefont {David}\
  \bibnamefont {Heckerman}},\ }\bibfield  {title} {\enquote {\bibinfo {title}
  {Correction for hidden confounders in the genetic analysis of gene
  expression},}\ }\href@noop {} {\bibfield  {journal} {\bibinfo  {journal}
  {Proceedings of the National Academy of Sciences}\ }\textbf {\bibinfo
  {volume} {107}},\ \bibinfo {pages} {16465--16470} (\bibinfo {year}
  {2010})}\BibitemShut {NoStop}%
\bibitem [{\citenamefont {Brem}\ and\ \citenamefont
  {Kruglyak}(2005)}]{brem2005landscape}%
  \BibitemOpen
  \bibfield  {author} {\bibinfo {author} {\bibfnamefont {Rachel~B}\
  \bibnamefont {Brem}}\ and\ \bibinfo {author} {\bibfnamefont {Leonid}\
  \bibnamefont {Kruglyak}},\ }\bibfield  {title} {\enquote {\bibinfo {title}
  {The landscape of genetic complexity across 5,700 gene expression traits in
  yeast},}\ }\href@noop {} {\bibfield  {journal} {\bibinfo  {journal} {PNAS}\
  }\textbf {\bibinfo {volume} {102}},\ \bibinfo {pages} {1572--1577} (\bibinfo
  {year} {2005})}\BibitemShut {NoStop}%
\bibitem [{\citenamefont {H{\"a}gg}\ \emph {et~al.}(2009)\citenamefont
  {H{\"a}gg}, \citenamefont {Skogsberg}, \citenamefont {Lundstr{\"o}m},
  \citenamefont {Noori}, \citenamefont {Nilsson}, \citenamefont {Zhong},
  \citenamefont {Maleki}, \citenamefont {Shang}, \citenamefont {Brinne},
  \citenamefont {Bradshaw}, \citenamefont {Bajic}, \citenamefont {Samnegard},
  \citenamefont {Silveira}, \citenamefont {Kaplan}, \citenamefont {Gigante},
  \citenamefont {Leander}, \citenamefont {de~Faire}, \citenamefont {Rosfors},
  \citenamefont {Lockowandt}, \citenamefont {Liska}, \citenamefont {Konrad},
  \citenamefont {Takolander}, \citenamefont {Franco-Cereceda}, \citenamefont
  {Schadt}, \citenamefont {Ivert}, \citenamefont {Hamsten}, \citenamefont
  {Tegner},\ and\ \citenamefont {Bj{\"o}rkegren}}]{hagg2009multi}%
  \BibitemOpen
  \bibfield  {author} {\bibinfo {author} {\bibfnamefont {S.}~\bibnamefont
  {H{\"a}gg}}, \bibinfo {author} {\bibfnamefont {J.}~\bibnamefont {Skogsberg}},
  \bibinfo {author} {\bibfnamefont {J.}~\bibnamefont {Lundstr{\"o}m}}, \bibinfo
  {author} {\bibfnamefont {P.}~\bibnamefont {Noori}}, \bibinfo {author}
  {\bibfnamefont {R.}~\bibnamefont {Nilsson}}, \bibinfo {author} {\bibfnamefont
  {H.}~\bibnamefont {Zhong}}, \bibinfo {author} {\bibfnamefont
  {S.}~\bibnamefont {Maleki}}, \bibinfo {author} {\bibfnamefont {M.~M.}\
  \bibnamefont {Shang}}, \bibinfo {author} {\bibfnamefont {B.}~\bibnamefont
  {Brinne}}, \bibinfo {author} {\bibfnamefont {M.}~\bibnamefont {Bradshaw}},
  \bibinfo {author} {\bibfnamefont {V.~B.}\ \bibnamefont {Bajic}}, \bibinfo
  {author} {\bibfnamefont {A.}~\bibnamefont {Samnegard}}, \bibinfo {author}
  {\bibfnamefont {A.}~\bibnamefont {Silveira}}, \bibinfo {author}
  {\bibfnamefont {L.~M.}\ \bibnamefont {Kaplan}}, \bibinfo {author}
  {\bibfnamefont {B.}~\bibnamefont {Gigante}}, \bibinfo {author} {\bibfnamefont
  {K.}~\bibnamefont {Leander}}, \bibinfo {author} {\bibfnamefont
  {U.}~\bibnamefont {de~Faire}}, \bibinfo {author} {\bibfnamefont
  {S.}~\bibnamefont {Rosfors}}, \bibinfo {author} {\bibfnamefont
  {U.}~\bibnamefont {Lockowandt}}, \bibinfo {author} {\bibfnamefont
  {J.}~\bibnamefont {Liska}}, \bibinfo {author} {\bibfnamefont
  {P.}~\bibnamefont {Konrad}}, \bibinfo {author} {\bibfnamefont
  {R.}~\bibnamefont {Takolander}}, \bibinfo {author} {\bibfnamefont
  {A.}~\bibnamefont {Franco-Cereceda}}, \bibinfo {author} {\bibfnamefont
  {E.~E.}\ \bibnamefont {Schadt}}, \bibinfo {author} {\bibfnamefont
  {T.}~\bibnamefont {Ivert}}, \bibinfo {author} {\bibfnamefont
  {A.}~\bibnamefont {Hamsten}}, \bibinfo {author} {\bibfnamefont
  {J.}~\bibnamefont {Tegner}}, \ and\ \bibinfo {author} {\bibfnamefont
  {J.}~\bibnamefont {Bj{\"o}rkegren}},\ }\bibfield  {title} {\enquote {\bibinfo
  {title} {{{M}ulti-organ expression profiling uncovers a gene module in
  coronary artery disease involving transendothelial migration of leukocytes
  and {L}{I}{M} domain binding 2: the {S}tockholm {A}therosclerosis {G}ene
  {E}xpression ({S}{T}{A}{G}{E}) study}},}\ }\href@noop {} {\bibfield
  {journal} {\bibinfo  {journal} {PLoS Genetics}\ }\textbf {\bibinfo {volume}
  {5}},\ \bibinfo {pages} {e1000754} (\bibinfo {year} {2009})}\BibitemShut
  {NoStop}%
\end{thebibliography}

%

\end{document}